\documentclass[aps,prl,twocolumn,showpacs]{revtex4}
\usepackage{graphicx}
\usepackage{color}
\usepackage{amssymb}
\usepackage{amsmath}

\newcommand{\PRLsection}[1]{\noindent {\it#1} -}

\makeatletter

\begin{document}

\title{Atom-Resonant Heralded Single Photons by ``Interaction-Free Measurement''}

\author{Florian Wolfgramm, Yannick A. de Icaza Astiz, Federica A. Beduini, Alessandro Cer\`{e}, and Morgan W. Mitchell}
\affiliation{ICFO - Institut de Ciencies Fotoniques, Mediterranean
Technology Park, 08860 Castelldefels (Barcelona), Spain}

\date{1 February 2010}

\begin{abstract}
We demonstrate the generation of rubidium-resonant heralded single
photons for quantum memories. Photon pairs are created by
cavity-enhanced down-conversion and narrowed in bandwidth to 7~MHz
with a novel atom-based filter operating by ``interaction-free
measurement'' principles. At least 94\% of the heralded photons
are atom-resonant as demonstrated by a direct absorption
measurement with rubidium vapor. A heralded auto-correlation
measurement shows $g_c^{(2)}(0)=0.040 \pm 0.012$, i.e.,
suppression of multi-photon contributions by a factor of 25
relative to a coherent state. The generated heralded photons can
readily be used in quantum memories and quantum networks.
\end{abstract}


\maketitle

\begin{figure}[t]
\includegraphics[width=0.4\textwidth]{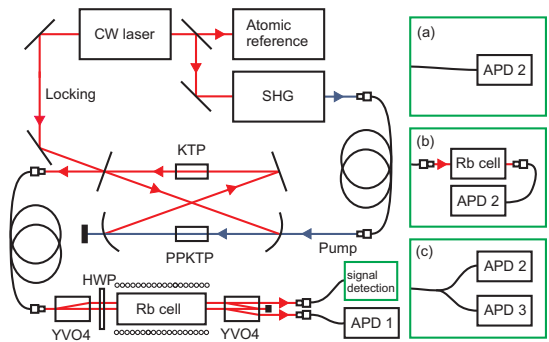}
\caption{Experimental setup. SHG, second harmonic generation
cavity; PPKTP, phase-matched nonlinear crystal; KTP, compensating
crystal; YVO4, Yttrium Vanadate crystal; HWP, half wave plate;
APD, avalanche photo-diode. (a)-(c), different measurement
scenarios for the signal photon detection. \label{img:Setup}}
\end{figure}

\PRLsection{Introduction.} The availability of single photons is a
crucial requirement in quantum information, quantum communication
and quantum metrology. For quantum networks, the photons
(\emph{flying qubits}) should be resonant with atoms
(\emph{stationary qubits}) for storage and/or processing. For this
reason, it has been an important goal of quantum optics to produce
high-purity single photons capable of interaction with atoms.
While there exist a number of different single-photon sources,
most of these do not fulfill all necessary requirements
\footnote{A review of narrow-band generation methods and their
limitations is provided in Reference
\cite{Neergaard-Nielsen2007}}. The most widely used heralded
single-photon source, spontaneous parametric down-conversion
(SPDC) \cite{Grangier1986,Fasel2004}, produces photons with a
spectral width orders of magnitude larger than typical atomic
natural linewidths. Passive filtering of SPDC photons is possible
and has been demonstrated \cite{Piro2010}, but shows low count
rates that are not sufficient for many tasks. Cavity-enhancement
of the down-conversion process has established itself in recent
years as a method to not only enhance the total photon rate, but
at the same time to enhance the emission into the spectral and
spatial modes of the cavity, producing high-purity photon states
at high rates
\cite{Ou1999,Kuklewicz2006,Neergaard-Nielsen2006,Neergaard-Nielsen2007,
Scholz2007,Wolfgramm2008,Bao2008,Scholz2009}.
\\
Bocquillon et al. \cite{Bocquillon2009} identify two critical
figures of merit for heralded single-photon sources. The first,
$g_{S,I}^{(2)}(\tau)$, describes the cross-correlation of signal
and idler beams, a measure of reliability of the heralding
mechanism. The second, $g_c^{(2)}(\tau)$, describes the
conditional auto-correlation of the signal beam, a measure of the
single-photon character of the heralded state. $g_c^{(2)}(0)<1$
indicates non-classical behavior; $g_c^{(2)}(0)=0$ for an ideal
source.
\\
Experiments using a cavity-enhanced, but unfiltered, source
\cite{Scholz2009} have demonstrated $g^{(2)}(0)<1$, but work in a
regime where many longitudinal frequency modes, spread over tens
of GHz, contribute to the signal. A cavity-enhanced source with
optical-cavity filtering of the heralding (idler) beam and
homodyne detection of the signal produced highly non-classical
states: 70\% of the heralded pulses contained a single photon in
the mode to which the detection was sensitive
\cite{Neergaard-Nielsen2006,Neergaard-Nielsen2007}. Undetected
modes, however, contained photons spread over a large bandwidth. A
recent experiment reports nearly 10\% efficient atom-storage of
beams from filtered cavity-enhanced SPDC, implying at least 10\%
atom-resonance, but made no measurement of $g_c^{(2)}(0)$
\cite{Jin2010}. To date, no SPDC single-photon source has
demonstrated atom-resonance of more than a small fraction of its
output.
\\
Here we demonstrate the generation of atom-resonant heralded
single photons with high spectral purity: within the detection
window of 400-1000~nm (450~THz), at least $94\%$ of the photons
are in a single, 7~MHz-bandwidth mode at the D$_1$ line of
$^{87}$Rb. Multi-photon contamination is at most $4\%$. We achieve
this using an atom-based filter, inspired by the
``interaction-free measurement'' (IFM) strategy of Elitzur and
Vaidman \cite{Elitzur1993} (also known as ``quantum
interrogation'' \cite{Kwiat1999}). The IFM proposal is based on a
balanced Mach-Zehnder interferometer in which due to destructive
interference one of the two output ports is dark. The presence of
an opaque object in either interferometer arm changes the
interference and thus increases the probability of a photon
exiting through the dark port. In our filtering scheme the object
is a hot atomic vapor which is opaque at the transition
frequencies. This guarantees the frequency of photons exiting
through the dark port to be at an atomic transition.
\\
IFM experiments have been proposed and demonstrated in different
systems \cite{Paraoanu2006} and for a variety of applications
including imaging \cite{White1998} and quantum computing
\cite{Hosten2006,Mitchison2007,Vaidman2007}.
\\
Intrinsic stability and intrinsic atom-resonance make our system
robust and attractive for quantum networking applications. Our IFM
filtering technique could be used also with solid-state ensembles
\cite{Riedmatten2008,Hedges2010}.
\\
\PRLsection{Experimental setup.} The experiment combines a
cavity-enhanced down-conversion source locked to a rubidium
transition, described in detail in \cite{Wolfgramm2008,
Wolfgramm2010} and an intrinsically atom-resonant narrow-band
filter, described in \cite{Cere2009}. The setup is shown
schematically in Fig.~\ref{img:Setup}.
\\
A single-frequency diode laser is locked to the
5$^{2}$S$_{1/2}$(F=2)$\rightarrow$5$^{2}$P$_{1/2}$(F'=1)
transition of the D$_1$ line of $^{87}$Rb. Part of the laser
output is frequency doubled and pumps the cavity-enhanced
down-conversion system at a typical pump power of 25~mW. Type-II
phase-matched down conversion takes place in a 2~cm-long
phase-matched periodically poled potassium titanyl phosphate
(PPKTP) crystal. Another KTP crystal (neither pumped nor
phase-matched) inside the cavity is temperature tuned to achieve
simultaneous resonance of signal and idler modes. The type-II
process generates photon pairs with mutually perpendicular
polarizations. This allows for straightforward separation of
signal and idler photon and also for easy generation of
polarization entanglement. A locking beam from the same diode
laser, and therefore also at the same rubidium transition
frequency, is used to stabilize the cavity length. In this way, we
guarantee the presence of frequency-degenerate cavity modes at the
atomic transition frequency. After leaving the cavity, the
generated photon pairs are coupled into a single-mode fiber.

The pumped, nonlinear cavity acts as a sub-threshold optical
parametric oscillator and generates resonant pairs of modes. With
the cavity locked and doubly-resonant for signal and idler modes,
the output spectrum is determined by the 148~GHz phase-matching
bandwidth of the down-conversion process. Within this envelope the
spectrum consists of hundreds of non-degenerate frequency modes
spaced by the free spectral range of 490~MHz, centered around the
degenerate mode at the rubidium transition frequency.

To achieve filtering that guarantees a high ratio between
degenerate and non-degenerate modes, e.g., a signal-to-noise ratio
of 90\%, requires an extinction ratio of several thousand, over a
bandwidth of hundreds of GHz. In principle this filtering can be
achieved with optical cavities. Consecutive cavities with
incommensurate free spectral ranges have been used in other
experiments
\cite{Neergaard-Nielsen2006,Neergaard-Nielsen2007,Piro2010}, but
do not appear to reach high rejection ratios. Neergaard-Nielsen et
al. report a 20\% discrepancy in effective signal detection
efficiency, and give ``insufficient suppression of uncorrelated
frequency modes in the series of trigger filters'' as a likely
explanation \cite{Neergaard-Nielsen2007}. A small misalignment or
aberration would be sufficient to couple into higher modes and
spoil the extinction ratio.

In contrast, our filter operates by principles of
``interaction-free measurement'' \cite{Elitzur1993,Kwiat1995a} and
combines extremely broadband optics (birefringent polarizers) with
extremely narrow-band optics (atoms) with a large angular
acceptance, thus practically insensitive to mode misalignment.

As shown in Fig.~\ref{img:Setup}, a YVO$_4$ crystal separates
horizontally and vertically polarized photons by 1 mm. The
polarization modes travel parallel to each other through a hot
rubidium cell of isotopically pure $^{87}$Rb, optically pumped by
a single-frequency laser resonant to the F=2$\rightarrow$F'=3
transition of the D$_2$ line of $^{87}$Rb (not shown). Due to
Doppler shifts, the optical pumping only effects a portion of the
thermal velocity distribution, and creates a circular dichroism
with a sub-Doppler linewidth of about 80~MHz. A second YVO$_4$
crystal introduces a second relative displacement, which can
re-combine or further separate the photons, depending on
polarization. Separated photons are collected, while re-combined
photons are blocked. A half wave plate is used to switch between
the ``active'' configuration, in which only photons that change
polarization in the cell are collected, and the ``inactive''
configuration, in which photons that do not change are collected.
In the ``active'' configuration, the system acts as an IFM
detector for polarized atoms: a photon is collected only if it
experiences a polarization change, i.e., if it is resonant with
the optically pumped atoms, which absorb one circular component of
the photon polarization state. Neighboring modes of the degenerate
mode at the rubidium transition are already 490~MHz detuned and
therefore outside of the filter linewidth of 80~MHz. The
out-of-band extinction ratio is $\geq$35~dB. The filter
transmission is optimized by adjusting the overlap between pump
and single-photon mode, the rubidium vapor temperature and the
magnitude of a small orienting applied magnetic field. The
temperature is set to 65$^{\circ}$C, which corresponds to an
atomic density of $5\cdot 10^{11}$ cm$^{-3}$. The measured filter
transmission of 10.0\% for horizontal polarization and 9.5\% for
vertical polarization is limited by pump power and in principle
can reach 25\% \cite{Cere2009}. To avoid contamination of the
single-photon mode by scattered pump light, the pump enters the
vapor cell at a small angle and counter-propagating to the
single-photon mode. Interference filters centered on 795~nm
further reject the 780~nm pump light with an extinction ratio of
$>$$10^{5}$. The measured contribution from pump photons is below
the detectors' dark count rate. Each output is coupled into
single-mode fiber. One is detected directly on a fiber-coupled
avalanche photo diode (APD, Perkin Elmer SPCM-AQ4C). The other is
used for subsequent experiments. Photon detections are recorded by
a counting board (FAST ComTec P7888) for later analysis.
\\
\PRLsection{Time-correlation measurements.} First, the time
distribution of the difference in arrival time between signal and
idler photons is analyzed in absence of the filter
(Fig.~\ref{img:Setup}(a)). We follow the theory developed in
\cite{Fasel2004,Herzog2008,Scholz2009,Bocquillon2009}. The
cross-correlation function between signal and idler modes is
\begin{equation}
g_{S,I}^{(2)}(\tau) \equiv \frac{\langle E_S^{\dag}(t+\tau)
E_I^{\dag}(t) E_I(t) E_S(t+\tau) \rangle}{\langle E_I^{\dag}(t)
E_I(t) \rangle
\langle E_S^{\dag}(t+\tau) E_S(t+\tau) \rangle}, \\
\end{equation}
where $E_{S,I}$ are the operators of the signal and idler fields.
In the case of doubly-resonant cavity-enhanced down-conversion it
takes this form:
\begin{equation}
    \begin{split}
        g_{S,I}^{(2)}(\tau) \propto {} & \Bigg| \sum_{m_S, m_I = 0}^\infty \frac{\sqrt{\gamma_S \, \gamma_I \, \omega_S \, \omega_I}}{\Gamma_S + \Gamma_I} \\
                        & \times
                      \begin{cases}
                                e^{-2 \pi \Gamma_S (\tau-(\tau_0/2))}{\rm sinc}{(i \pi \tau_0 \Gamma_S)} \hspace{3 mm} \hspace{0.5 mm} \tau \geqslant \frac{\tau_0}{2}\\
                                e^{+2 \pi \Gamma_I (\tau-(\tau_0/2))}{\rm sinc}{(i \pi \tau_0 \Gamma_I)} \hspace{4 mm} \hspace{0.5 mm} \tau < \frac{\tau_0}{2}
                      \end{cases} \hspace{-4 mm} \Bigg|^2,
    \end{split}
    \label{eq:cross}
\end{equation}
where $\gamma_{S,I}$ are the cavity damping rates for signal ($S$)
and idler ($I$), $\omega_{S,I}$ are the central frequencies,
$\tau_0$ is difference between the transit times of a signal and
idler photon through the SPDC crystal, $\Gamma_{S,I} \equiv
\gamma_{S,I}/2+i m_{S,I} \Delta \omega_{S,I}$ with mode indices
$m_{S,I}$ and free spectral ranges $\Delta \omega_{S,I}$
\cite{Scholz2009,Herzog2008}. Due to compensation, $\Delta\omega_S
= \Delta\omega_I \equiv \Delta\omega$ in our cavity.

\begin{figure}[b]
\centering
\includegraphics[width=0.4\textwidth]{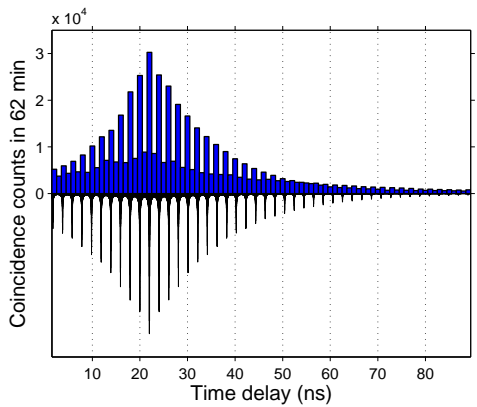}
\caption{Arrival time histogram of unfiltered photon pairs;
experimental data (upper bars) and theory (lower bars). The
frequency-comb structure is reflected by a comb-like structure in
the temporal domain. The visibility of the experimental data is
limited by time resolution of the counting electronics.
\label{img:Unfiltered}}
\end{figure}

\begin{figure}[b]
\centering
\includegraphics[width=0.4\textwidth]{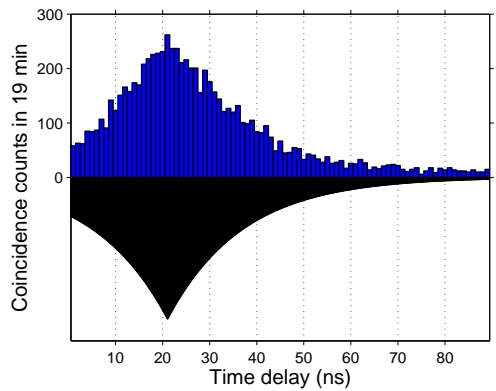}
\caption{Arrival time histogram of filtered photon pairs;
experimental data (upper bars) and theory (lower bars). The
disappearance of the comb structure in the filtered case indicates
the single-mode character of the filtered fields.
\label{img:Filtered}}
\end{figure}

We first measure the $g_{S,I}^{(2)}(\tau)$-function with the
filter in the ``inactive'' configuration at a much reduced pump
power. The histogram of the difference in arrival time between
detection events in the two APDs is shown in
Fig.~\ref{img:Unfiltered}.
The blue bars represent the coincidence event detections within
time bins of 1~ns, the resolution of the counting board. The black
bars, drawn inverted for better visibility, show the theoretical
prediction based on Eq.~(\ref{eq:cross}). The height of the theory
histogram, the only free parameter, has been set to match the
height of the data. Experimental and theoretical results are in
excellent agreement. The comb-like structure of the histogram is a
consequence of interference between different frequency modes. The
temporal spacing between neighboring peaks corresponds to the
cavity round-trip time $1/\Delta\omega \approx$ 2.04~ns.
\\
When the filter is ``active'', the arrival time difference
histogram shows a smooth double-exponential shape, without
multi-mode interference (Fig.~\ref{img:Filtered}).
This already indicates that only a single frequency mode is
transmitted through the filter. The theory (lower black bars) is
given by Eq.~(\ref{eq:cross}) for a single mode
($\Gamma_{S,I}$=$\gamma_{S,I}/2$). The data shows a very low
background noise level. Throughout, raw data are shown; background
coincidences have not been subtracted.
\\
In this experiment we are interested in time correlations, but it
is interesting to ask if other kinds of correlations and possible
entanglement, e.g. in polarization or in frequency, are also
preserved by the filter. By design, the filter should transmit
nearly equally different frequency and polarization components of
the selected cavity mode, preserving correlations: absorptive and
refractive effects vary on the scale of the 80~MHz absorption
linewidth, large relative to the 7~MHz of the cavity mode. Also,
the axial magnetic field scrambles any linear birefringence or
dichroism, giving equal response for the two linear polarizations.
Preliminary results indicate that the degree of polarization as
well as the entanglement in a polarization entangled state are not
changed significantly by the filter. A detailed study of this will
be the subject of a future publication.

\begin{figure}[t]
\centering
\includegraphics[width=0.4\textwidth]{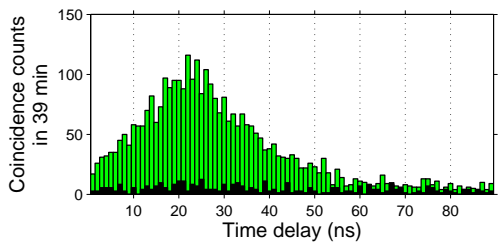}
\caption{Arrival time histogram of filtered photon pairs after
passing the signal photons through a rubidium vapor cell at an
optical density of 0.3 (upper green bars) and at an optical
density of 6 (lower black bars). \label{img:Cell}}
\end{figure}

\PRLsection{Atom-resonance.} To measure the atom-resonant
fraction, we let the filtered photons of the signal arm propagate
through a rubidium vapor cell (Fig.~\ref{img:Setup}(b)). At room
temperature, the cell's optical density (OD) is low (0.3)
corresponding to a transmission of 74\% and coincidences between
the detection events on the two APDs are observed
(Fig.~\ref{img:Cell}, upper green bars). By heating the rubidium
cell, an optical density of 6, or 0.25\% resonant transmission, is
reached. The coincidences drop to the background level
(Fig.~\ref{img:Cell}, lower black bars).
Within a coincidence window of 40~ns, the ratio of raw OD 0.3
coincidences to raw OD 6 coincidences is 11.6:1, indicating
rubidium resonance of at least 94\% of the photons.

\PRLsection{Suppression of multi-photon events.} The signal
auto-correlation function, given a trigger detection of the idler,
is \cite{Fasel2004, Bocquillon2009, Scholz2009}
\begin{equation}
g_c^{(2)}(\tau)=\frac{\langle E_S^{\dag}(t+\tau) E_S^{\dag}(t)
E_S(t) E_S(t+\tau) \rangle}{\langle E_S^{\dag}(t) E_S(t) \rangle
\langle E_S^{\dag}(t+\tau) E_S(t+\tau) \rangle}. \label{eq:auto}
\end{equation}
The crucial figure of merit is the value of the auto-correlation
function {of signal photons} $g_c^{(2)}(\tau)$ at $\tau=0$. We
measure $g_c^{(2)}(0)$ as follows: the signal mode is split by a
50/50 beam splitting fiber and the coincidences between the idler
detector (APD1) and the two signal detectors (APD2 and APD3) are
analyzed (Fig.~\ref{img:Setup}(c)). The detection of an idler
photon defines a coincidence window of 40~ns, symmetrical around
the detection time. Individual and coincident detections in this
time window give singles  counts $N_2, N_3$, while detections at
both APD2 and APD3 give the coincidence count $N_{23}$. $N_{23}$
corresponds to unwanted multi-photon contributions which are very
low in our experiment. To accurately estimate $N_{23}$, we measure
for large coincidence windows of up to 2000~ns, extrapolate down
to 40~ns, and multiply by two, to account for possible bunching
\cite{Bocquillon2009,Scholz2009}. We then calculate
\begin{equation}
g_c^{(2)}(0) \approx \frac{N_{23}N_1}{N_2 N_3},
\end{equation}
where $N_1$ is the number of idler trigger events
\cite{Grangier1986,Fasel2004,Bocquillon2009}. We note that this
gives an upper limit for $g_c^{(2)}(0)$, due to the conservative
bunching factor and the finite time window. We find
$g_c^{(2)}(0)\le 0.040 \pm 0.012$, 80 standard deviations below
the classical limit of 1.

\PRLsection{Summary.} Using an ultra-bright cavity-enhanced
down-conversion source and an atom-based filter operating by
``interaction-free measurement'' principles, we have generated for
the first time narrow-band, high-spectral purity, atom-resonant
heralded single photons from SPDC. Of the generated photons, 94\%
are resonant to a rubidium transition frequency. A
$g_c^{(2)}$-measurement shows an upper limit of
$g_c^{(2)}(0)=0.040\pm0.012$ corresponding to a reduction of
multiple photon events by a factor of at least 25 compared to a
coherent state. The source is an ideal tool for atom-photon
interactions at the single-photon level, for quantum memories in
EIT media \cite{Eisaman2005} and solid-state systems
\cite{Riedmatten2008,Hedges2010} and single-photon single-atom
interfaces \cite{Tey2008,Piro2010}.
\begin{acknowledgments}
We acknowledge useful discussions with P. Kwiat, H. de Riedmatten
and C. Vitelli. This work was supported by the Spanish Ministry of
Science and Innovation under the Consolider-Ingenio 2010 Project
``Quantum Optical Information Technologies'' and the ILUMA project
(No. FIS2008-01051) and by an ICFO-OCE collaborative research
program. F.~W. is supported by the Commission for Universities and
Research of the Department of Innovation, Universities and
Enterprises of the Catalan Government and the European Social
Fund.
\end{acknowledgments}


\end{document}